# Disorder induced loss of magnetization

# in Lieb's graphene quantum dots

B. Jaworowski*, P. Potasz, A. Wójs

*Institute of Physics, Wroclaw University of Technology, Wroclaw, Poland;*

**\*** Corresponding author: jmvarelse@gmail.com

Tel: +98-9133452131

## Abstract

The stability of the magnetization in Lieb graphene quantum dots (GQD) against disorder is studied. Such systems exhibit the degenerate shell of edge states in the middle of the energy gap occupied by spin polarized electrons. Disorder affects the energy spectrum and leads to the removal of the degeneracy, and in a consequence the loss of the magnetization. We find that there is a critical value of disorder strength over which the magnetization starts dropping. We first consider GQDs with different shapes and edge termination and compare the effect of bulk and edge disorder. We find that bulk disorder influences the energy spectrum far away from the Fermi level while edge disorder affects states in the vicinity of it. We next focus on Lieb GQDs with a single zigzag edge. The stability of the ferromagnetic order against disorder strength is analyzed for structures with a different edge length. The probability of getting maximal spin polarization is determined.

# 1. Introduction

Since its exfoliation in 2004 [1], graphene continuously astounds the scientific community from point of view of pure theory as well as possible applications [2-6]. In graphene, low energy quasi-particles behave like massless Dirac fermions what gives opportunity to study relativistic effects [2, 3]. For applications in electronics, particularly promising are the very high mobility of carrries in graphene, optical transparency, good quality of crystal and planar structure [4-6]. The lack of the band gap in the band structure of graphene limits these applications. There are several proposition of creating the energy gap in graphene [6]. One possibility is to create graphene nanostructures by carving graphene into desired shapes [7-19]. The energy gap opens due to quantum confinement effect. A variety of graphene nanostructures have yet been studied, including graphene nanoribbons (GNR) [8, 9], graphene quantum dots (GQD) and rings (GQR) with different shapes: triangular, hexagonal, rectangular, and randomly shaped [10-22]. Independently of the shape, the energy gap opens. It was shown that not only shape but also edge termination determines the electronic properties of graphene nanostructures. In particular, zigzag type of edge leads to an appearance of edge states in a vicinity of the Fermi energy, and structures can reveal the magnetic order due to the polarization of spins of electrons occupying these states [8, 9, 13-17]. The magnetization is related to the presence of the degenerate shell of edge states in the middle of the energy gap. The degeneracy of the shell is determined by an imbalance between numbers of atoms belonging to two sublattices of a graphene bipartite honeycomb lattice. According to Liebs theorem regarding the total spin of the ground state in the Hubbard model for a bipartite lattice system with imbalanced number of two types of atoms, $S=(N_A-N_B)/2$, where $N_A$ and $N_B$ are the numbers of two types of atoms. This type of GQDs can be called Lieb's GQD.

In any real material, the perfect lattice is altered by disorder. Among disorder types in graphene-based structures one can list vacancies, adsorbates, impurities, topological lattice defects (e.g. pentagons and heptagons), and randomly removed atoms from the edge - edge roughness due to problems with a fabrication with atomic precision. Also, graphene sheets usually are not perfectly flat; a tendency to stabilize the structure by forming ripples is observed [23]. Moreover, additional sources of disorder appear when graphene sheet is deposited on a substrate. An interaction with a substrate causes displacement of atoms [24]. The charge traps and charged impurities in a substrate are the source of random electrostatic potentials. Experiments show that in epitaxial graphene short range defects are present, while there are very rare in suspended graphene [25].

Disorder in graphene can be modeled in many ways, including changes of values of the hopping integral in a tight-binding model corresponding to a change of bond lengths [24], using Anderson model to study short-range scattering centers [26], using long-range potentials to model charged impurities [27, 28], or introducing random vector potentials corresponding to ripples [29]. It is expected that the effect of disorder is stronger close to edges comparing to the center of nanostructures, thus different values for edge and bulk disorder are usually taken. Methods allowing investigating the effect of disorder are described in several review articles, e. g. see Ref. 7, 30-32.

The effect of disorder on electronic properties of graphene nanostructures was intensively studied [33-39]. For graphene nanoribbons, the research was done from perspective of transport properties: the conductivity and mean free path was calculated [33]. Disorder was investigated using Anderson model in a bulk form. Changes of the density of states (DOS) and localization length of wavefunctions were analyzed. DOS were mostly affected around van Hove singularities and in the vicinity of the Fermi energy where peaks corresponding to edge states appear. It was taken into account that edges may be more prone to disorder. Edge disorder was modeled in two ways: using Anderson disorder placed only at the edge lattice sites or randomly removing selected atoms from the edge. Both models lead to quasi-one-dimensional Anderson localization [33]. The effect of disorder was also studied in graphene quantum dots. The energy level statistics, the transition to quantum chaos, and transport properties were investigated [34-36]. The stability of edge states in a presence of edge roughness in large GQD, consisting of up to 75 000 atoms, was considered [37, 38]. The effect of bulk and edge Anderson disorder, and random magnetic fluxes on electronic and transport properties in electrostatically induced circular graphene quantum dots were investigated. The peaks of density of states, which correspond to bound states remained sharp only in a case of edge Anderson disorder, and were broadened and disappeared when the disorder strength was sufficiently high [39].

Disorder affects the energy spectra of GQD, thus also the degenerate shell in Lieb's GQDs. Because of possible removal of the degeneracy, an important question arises regarding the influence of disorder on the magnetic moment. In this work, we analyze the stability of the magnetization of Lieb GQDs as a function of the short-range disorder strength using Anderson model within tight-binding and mean-field Hubbard models. We first investigate the effect of disorder on the density of states for GQDs with different shapes and edge termination. A comparison between bulk and edge disorder is presented. We next focus on Lieb GQDs with a single zigzag edge. We analyze the stability of the magnetization as a function of the disorder strength. We consider structures with different zigzag edge lengths and determine the probability of finding the ground state with a given spin. The critical value over which the spin polarization drops is determined.

## 2. Method

The electronic properties of GQDs within a single-particle approximation can be investigated using tight-binding model and the magnetic properties of Lieb GQD can be described using Hubbard model in a mean-field approximation. The results obtained by above methods are in good agreement with Density Functional Theory calculations [13, 21], assuming that the edge reconstruction is not present. The edge reconstruction in triangular Lieb GQDs was shown to destroy the magnetization due to an increase of the width of the degenerate shell [21]. Tight-binding model is described be following Hamiltonian

$$H_0 = \sum_{i\sigma} \varepsilon_i c_{i\sigma}^+ c_{i\sigma} - \sum_{<i,j>\sigma} t c_{i\sigma}^+ c_{j\sigma} - \sum_{<<i,j>>\sigma} t_2 c_{i\sigma}^+ c_{j\sigma} , \qquad (1)$$

where $c_{i\sigma}^+$ ($c_{i\sigma}$) are a creation (annihilation) operator of an electron at lattice site $i$ with spin $\sigma$, $\varepsilon_i$ is site-dependent on-site kinetic energy, $t$ and $t_2$ are hopping integrals to the nearest and next nearest neighbors, respectively. In our calculations, for systems in the absence of disorder following values are taken: $t = 2.84$ eV, $t_2 = 0.1$ eV, $\varepsilon_i = 0$. Disorder is modeled in the

Anderson form taking $\varepsilon_i$ from a box distribution, a uniform probability from the range [−W/2, W/2]. The effect of bulk and edge disorder on energy spectra is analyzed for a disorder strength W = 2eV (W/t ≈ 0.7). In order to investigate the effect of disorder independent of the specific realization, we create 5000 realizations of Hamiltonian given by Eq. 1 with random values of on-site energies. The final DOS is obtained by averaging energy spectra over these 5000 realizations. We compare edge and bulk disorder, where $\varepsilon_i$ is chosen randomly with the same probability only atoms in the vicinity of the edge and for all atoms in the structure, respectively. We compare the effect of edge disorder on GQDs with different shape and edge termination, and also study the effect of disorder on a single zigzag edge.

For calculations of magnetization, Hubbard model in a mean-field approximation is used, with Hamiltonian

$$H = H_0 + U\sum_i \hat{n}_{i\downarrow}\langle\hat{n}_{i\uparrow}\rangle + \hat{n}_{i\uparrow}\langle\hat{n}_{i\downarrow}\rangle, \qquad (2)$$

where $H_0$ is given by Eq. (1), $\hat{n}_{i\uparrow}$ is the electron number operator for spin up at *i-th* lattice site, and $U = 1.1t$. The average values of spin densities on sites are calculated summing over all occupied states. A self-consistent procedure is used and after convergence process, total energies corresponding to different total spins are found, and the ground state total spin is determined. For each value of the disorder strength 500 realizations of self-consistent calculations are performed, and the probability of finding the ground state with a given total spin is calculated.

### 3. Result

Before we make an analysis of the stability of the magnetization in Lieb GQDs, we first compare the effect of edge disorder and bulk disorder on electronic properties of GQDs. In Fig. 1, three graphene quantum dots with different shapes and edge termination consisting of around N~1000 atoms are considered: armchair hexagon with N=1014 atoms in Fig. 1(a), zigzag hexagon with N=1014 atoms in Fig. 1(b) and zigzag triangle with N=1021 atoms in Fig. 1(c).

We compare density of states (DOS) obtained by summing over discrete energy levels approximated by using Gaussian functions with a standard deviation $\sigma = 0.16$ eV. In the absence of disorder, DOS of all three structures look similarly to that for graphene, with characteristic van Hove singularities around energy $E = \pm t$ [6]. All systems have energy gaps around the Fermi energy due to size quantization effects but they are not visible in Fig. 1 due to an overlap of Gaussian functions used to plot DOS. Magnitudes of the energy gaps are: in armchair hexagon $E_{gap} = 0.57$ eV, zigzag hexagon $E_{gap} = 0.03$ eV, and zigzag triangle is about $E_{gap} = 0.54$ eV. For structures with zigzag edges instead of vanishing density of states at the Fermi energy $E_F = 0$, there is an extra peak related to the presence of edge states, see Fig. 1(a) and 1(b). The edge states are responsible for the smallest value of the energy gap in zigzag hexagon. The peak at the Fermi energy is higher for zigzag triangle comparing to zigzag hexagon. Zigzag triangle has non-balanced number of two types of atoms in a honeycomb graphene lattice, which leads to an appearance of zero-energy degenerate edge states. The number of zero-energy states in the degenerate shell equals to the sublattice imbalance [20]. We note that these states have energy perfectly $E = 0$ only in a tight-binding model within a nearest neighbors approximation, for a Hamiltonian given by Eq. (1) with a neglected last term. The degeneracy is slightly lifted when the next-neighboring terms are included. When the disorder is present in the system, DOS slightly changes for all considered structures. DOS for the armchair hexagon is strongly affected around the van Hove singularities only by bulk disorder, while the effect of edge disorder is almost invisible, see Fig. 1(a). For structures with zigzag edges, the effects of bulk and edge disorders are observed around the van Hove singularities and in a vicinity of the Fermi energy, see Fig. 1(b) and 1(c). The peaks around the Fermi energy related to the edge states are broadened, which is not observed for armchair hexagon because there are no states localized at the edges. While bulk disorder affects both van Hove singularities and vicinity of the Fermi energy, it is related to the fact that in bulk disorder also variations of onsite energies of edge atoms are included, and this is the reason of changes of DOS in a vicinity of the Fermi energy. We note that while both types of disorder influence mostly van Hove singularities - a bulk disorder effect, and the vicinity of the Fermi energy - an edge disorder effect, also other regions of the spectrum are affected, not shown here. This becomes visible for smaller half-width of the Gaussians used to plot DOS. The results for bulk disorder in these GQDs are similar to the results for bulk disorder in nanoribbons [33]. Above results show that in order to study the effect of disorder on low energy electronic properties, around Fermi energy, including the degeneracy responsible for the magnetic properties, only edge disorder has to be considered.

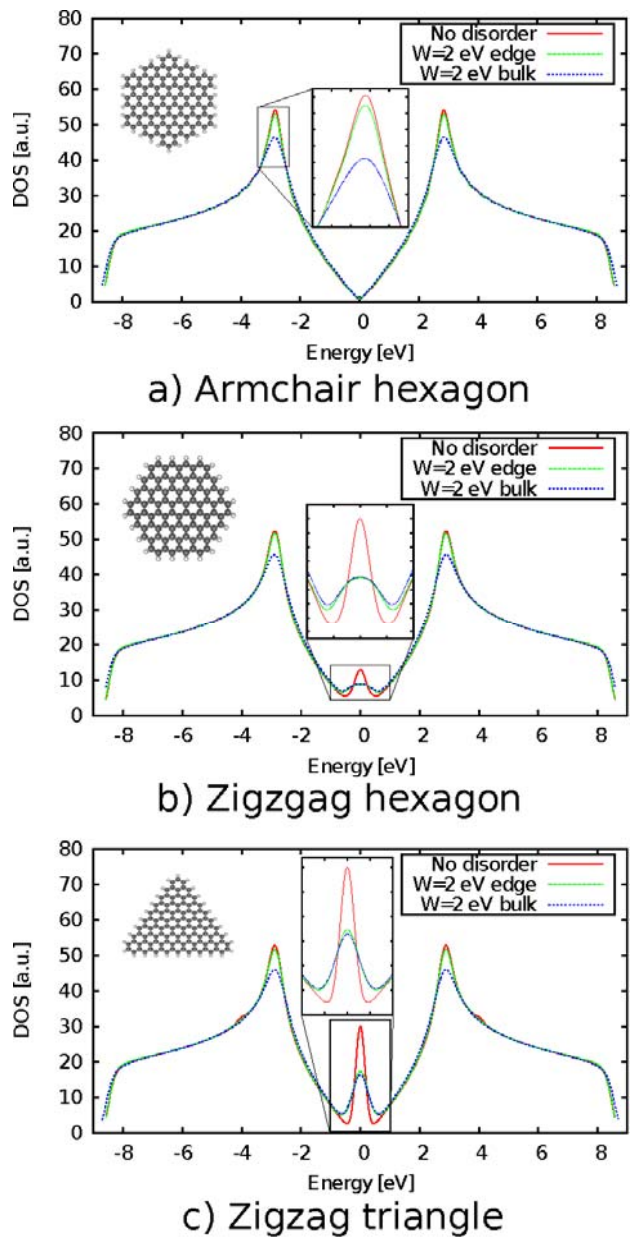

Fig. 1. Comparison of DOS under influence of bulk and edge disorder with W=2eV, W/t=0.7, for three GQD with different shapes and edge termination. In order to obtain DOS, each energy level is replaced by a Gaussian function with a standard deviation $\sigma = 0.16$ eV. For disordered GQD, the result is averaged over 5000 configurations. GQD structures in the insets are smaller dots of the same type in order to make edge details resolvable. The energy spectra around van Hove singularities in a) and the Fermi energy in b) and c) are enlarged.

In order to understand the role of disorder better, we investigate a structure with a single zigzag edge with different lengths. We have a freedom of choosing the shape and size of the considered structure with a given number of zero-energy states localized at a single zigzag edge, as long as following conditions are satisfied: i) a global imbalance between the number of two types of atoms is given – changing the imbalance would lead to increase or decrease of the number of zero-energy states. ii) there is only one zigzag edge – when more zigzag edges appear, the localization of the degenerate states around a single zigzag edge is not always ensured. We also note that, when the system contains large number of atoms, the energy gap closes and the degenerate shell of edge states is not well separated from the rest of the spectrum. Satisfying above conditions, we consider a pentagonal geometry with four armchair edges and one zigzag edge at the bottom. While such system can be difficult to fabricate, we consider it only in order to analyze disorder effects on a single zigzag edge, and our considerations should be valid for an entire class of Lieb GQDs. The lengths of armchair edges can change the total number of atoms, and in a consequence the magnitude of the energy gap, but can not influence the degenerate edge states localized on a zigzag edge. The number of the degenerate states is determined only by the length of the zigzag edge, what can be written as

$$N_{\text{deg}} = \frac{N_{edge} - 1}{3}, \qquad (3)$$

where $N_{edge}$ is the number of edge atoms on the zigzag edge. We note that when the left hand side of Eq. (3) gives noninteger value, more than one zigzag edge has to appear in the structure, and the degenerate states are not solely localized on one zigzag edge, but distributed over all zigzag edges. Thus, we concentrate on the lengths when $N_{edge} - 1$ is a multiply of 3. In Fig. 2 the tight-binding energy spectra obtained by diagonalizing Hamiltonian given by Eq. (1) are compared for the structure consisting of N=264 atoms and $N_{edge} = 19$ with and without disorder. According to Eq. (3), there are $N_{\text{deg}} = 6$ degenerate edge states. They are clearly separated from the rest of the spectrum by $E_{gap} = 0.65$ eV gap. Although the energy gap decreases with system size, for the structure containing N=966 atoms, $E_{gap} = 0.27$ so the degenerate shell is still well-separated from the rest of the spectrum. The disorder strength is chosen as W ≈ 0.94 eV (W/t =

0.33). It increases the dispersion of the degenerate states, but does not affect their total charge densities, obtained by summing over charge densities of all degenerate states, shown in the upper inset in Fig. 2, for the system with and without disorder - both densities are identical.

For the charge neutral system, all states below the Fermi level are doubly occupied by electrons with opposite spins, and the degenerate shell is half-filled. Using the mean-field Hubbard Hamiltonian given by Eq. (2) we obtain that a spin gap appears, shown in the lower inset in Fig. 2. All the degenerate states are filled by spin-down electrons and total spin of the ground state $S=3$. For non-disordered structure, this result is in agreement with Lieb's theorem [40], regarding the total spin of the ground state in the Hubbard model for a bipartite lattice system with imbalanced number of two types of atoms, $S=(N_A-N_B)/2$, where $N_A$ and $N_B$ are the numbers of two types of atoms and for considered system $N_A-N_B=6$. We want to examine the stability of the magnetic order against disorder on a single zigzag edge. The contribution to the magnetic moment comes from the degenerate states. All these states are localized near the zigzag edge. As long as the graphene quantum dot has only one zigzag edge, whose length is fixed, and the zero-energy shell is well separated from the rest of spectrum, magnetic properties are not changed when shape and size of GQD varies. When more zigzag edges appear in a structure, the degenerate states are distributed over all of them. We do not consider such situation here; however, we can expect that this can lead only to quantitative difference comparing to results presented in this work.

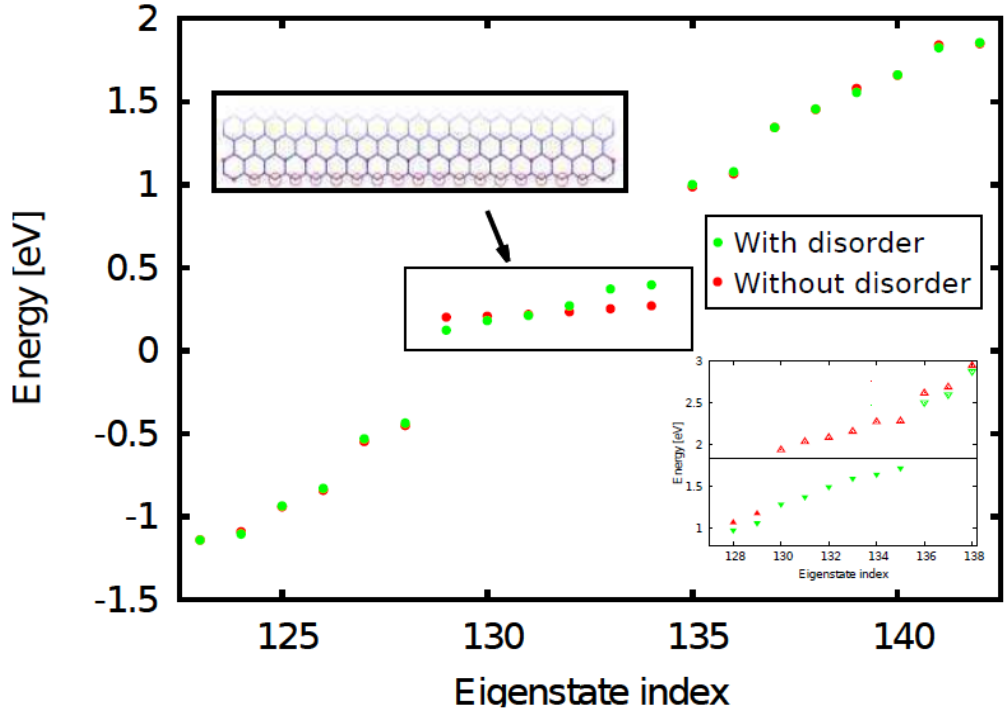

Fig. 2. Comparison of energy spectra for a pentagonal quantum dot with one zigzag edge of length $N_{edge}=19$ edge atoms (N=264 atoms in a whole structure) with and without disorder for one of the 500 studied realization of random on-site energies. The upper inset shows spin density for non-disordered and disordered dot for W/t=0.35 – both densities are almost identical. Lower inset shows Hubbard spectrum for the dot with disorder – without disorder the plot is qualitatively similar. The filled and empty triangles correspond to occupied and empty states. Red triangles are spin up states, while the blue ones are spin down states. Black line indicates the Fermi level.

We investigated the effect of disorder on four structures with different zigzag edge lengths, and in a consequence a number of the degenerate states from $N_{deg}$=3 to $N_{deg}$=6. For each value of disorder, 500 realizations of different disorder configurations are created, and each time total spin of the ground state is determined. The probability of getting a given total spin is obtained by averaging over all these realizations. The results are shown in Fig. 3 for the structure with $N_{deg}$=6 degenerate states in the energy spectrum. Without disorder, according to Lieb's theorem the total spin is S=3, that means all degenerate states are filled by spin polarized electrons. Increasing disorder up to value W/t=0.45 does not affect the spin polarization – maximal polarization of electrons from the degenerate shell occurs with a probability equal to 1. Around W/t=0.45 this probability start dropping, where the probability of getting spin S=2 as the ground state increases. Around W/t=0.76, the probabilities of maximal spin polarization S=3 and S=2 are equal. For further increase of the disorder, the ground state is expected to have total spin S=2, and also lower values of total spins, and the probability of getting maximal spin polarization vanishes. For structures with a different length of zigzag edge, similar behavior is observed. A maximal value of disorder strength to get the probability equal to 1 for maximal spin polarization oscillates around W/t ≈ 0.45, with W/t ≈ 0.5 for the system with $N_{deg}$ = 4, and W/t ≈ 0.39 for the system with $N_{deg}$ = 5. Thus, the ferromagnetic order is quite stable against disorder for all consider structures in a wide range of disorder strength, regardless of the edge length, while always at the end sufficiently strong disorder destroys the magnetization.

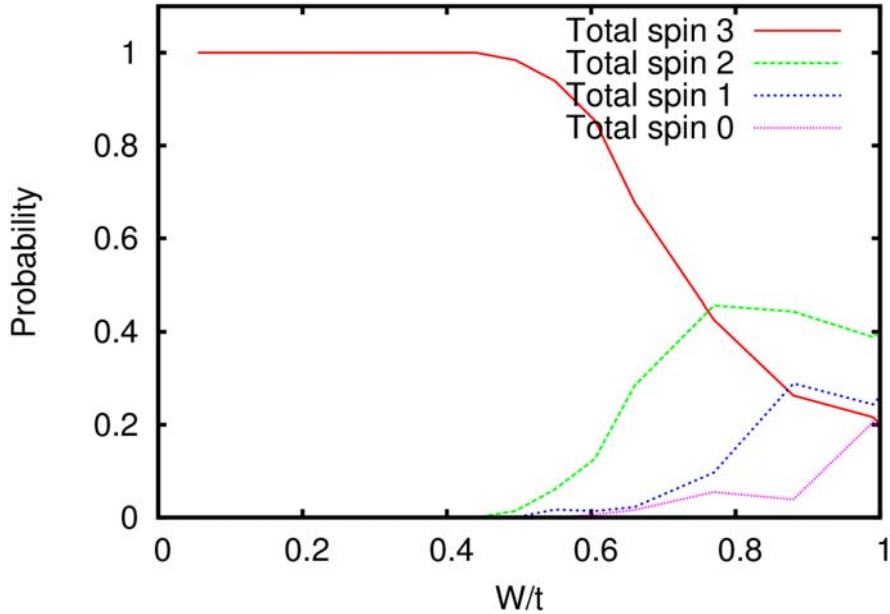

Fig. 3. A Probability that the ground state has a given spin, for a structure with a zigzag edge conisiting of $N_{edge} = 19$ atoms as a function of the disorder strength W/t. The results are obtained using mean-field Hubbard model, with 500 realizations of disorder for each W/t value.

We also compare the effect of bulk end edge disorder. Both types of disorder give similar results. For a system with $N_{deg}$=3, the disorder strength where the maximal polarization probability drops is similar, although the drop itself is slightly more abrupt. For $N_{deg}$=4, the critical strength is W/t≈0.45 in bulk case and W/ t≈0.5 in edge case. The differences are related to the fact that the edge states responsible for the magnetic order are not perfectly localized at the zigzag edge, therefore, although the main contribution come from the edge, disorder on other atoms would also influence it.

## 4. Conclusion

We investigated the stability of the magnetization in Lieb GQD against disorder. We analyzed structures with a single zigzag edge and different edge lengths. A large interval of disorder strength, below which the probability of getting maximal spin polarization is equal to 1, was shown. The loss of the magnetization for a sufficiently strong disorder in all studied systems was observed. We also compared the effect of bulk and edge disorder on DOS in GQDs. Bulk disorder changed DOS around van Hove singularities, while edge disorder influenced the vicinity of the Fermi energy where peaks corresponding to edge states are present.

## Acknowledgements

This work has been supported by the Polish NCN grant No. 2011/01/B/ST3/04/504 and by the Fellowship "MISTRZ" from the Foundation for Polish Science.

## References

[1] K. S. Novoselov, A. K. Geim, S. V. Morozov, D. Jiang, Y. Zhang, S. V. Dubonos, I. V. Grigorieva, and A. A. Firsov, Science 306, 666–669 (2004).

[2] K. S. Novoselov, A. K. Geim, S. V. Morozov, D. Jiang, M. I. Katsnelson, I. V. Grigorieva, S. V. Dubonos, and A. A. Firsov, Nature 438, 197 (2005).

[3] Y. Zhang, Y. W. Tan, H. L. Stormer, and P. Kim, Nature 438, 201 (2005).

[4] K. S. Novoselov and A. K. Geim, Nature 6, 183 (2007).

[5] M. I. Katsnelson, Materials Today 10, 20-27 (2007).

[6] A. H. Castro Neto, F. Guinea, N. M. R. Peres, K. S. Novoselov, and A. K. Geim, Rev. Mod. Phys. 81, 109–162 (2009).

[7] D. S. L. Abergel, V. Apalkov, J. Berashevich, K. Ziegler, and T. Chakraborty, Adv. Phys 59 261–482 (2010).

[8] Y.-W. Son, M. L. Cohen, and S. G. Louie, Phys. Rev. Lett. 97, 216803 (2006).

[9] Y.-W. Son, M. L. Cohen, and S.G. Louie, Nature 444 347-349 (2006).

[10] T. Yamamoto, T. Noguchi, and K. Watanabe, Phys. Rev. B 74, 121409 (2006).

[11] M. Ezawa, Phys. Rev. B 76, 245415 (2007).

[12] D. E. Jiang, B. G. Sumpter, and S. Dai, J. Chem. Phys. 127, 124703 (2007).

[13] J. Fernandez-Rossier, J. J. Palacios, Phys. Rev. Lett. 99, 177204 (2007).

[14] W. L. Wang, S. Meng, and E. Kaxiras, Nano Lett. 8, 241 (2008).

[15] W. L. Wang, O. V. Yazyev, S. Meng, and E. Kaxiras, Phys. Rev. Lett. 102, 157201 (2009).

[16] A. D. Güçlü, P. Potasz, O. Voznyy, M. Korkusinski, and P. Hawrylak, Phys. Rev. Lett. 103, 246805 (2009).

[17] P. Potasz, A. D. Güçlü, and P. Hawrylak, Phys. Rev. B 85, 075431 (2012).

[18] J. Akola, H. P. Heiskanen, and M. Manninen, Phys. Rev. B 77, 193410 (2008).

[19] D. A. Bahamon, A. L. C. Pereira, P. A. Schulz, Phys. Rev. B 79, 125414 (2009).

[20] P. Potasz, A. D. Guclu, and P. Hawrylak, Phys. Rev. B 82, 075425 (2010).

[21] P. Potasz, A. D. Guclu, O. Voznyy, J. A. Folk, and P. Hawrylak, Phys. Rev. B 83, 174441 (2011); O. Voznyy, A. D. Guclu, P. Potasz, and P. Hawrylak, Phys. Rev. B 83, 165417 (2011).

[22] Z. Z. Zhang, K. Chang, F. M. Peeters, Phys. Rev. B 77, 235411(2008).

[23] W. Bao, F. Miao, Z. Chen, H. Zhang, W. Jang, C. Dames, and C. N. Lau, Nature Nanotech 4, 562 (2004).

[24] D. Areshkin, D. Gunlycke, and C.T. White, Nano Lett. 7, 204 (2007).

[25] G. M. Rutter, J. N. Crain, N. P. Guisinger, T. Li, P. N. First, J. A. Stroscio, Science 317, 219-222 (2007).

[26] M. Amini, S. A. Jafari, and F. Shahbazi, Europhys. Lett. 87, 37002 (2009).

[27] K. Nomura, A.H. MacDonald, Phys. Rev. Lett. 98, 076602 (2007).

[28] Y.Y. Zhang, J. Hu, B. A. Bernevig, X. R. Wang, X. C. Xie, and W. M. Liu, Phys. Rev. Lett. 102, 106401 (2009).

[29] S.V. Morozov, K.S. Novoselov, M.I. Katsnelson, F. Schedin, L.A. Ponomarenko, D. Jiang, A.K. Geim, Phys. Rev. Lett. 97, 016801(2006).

[30] E. R. Mucciolo and C. H. Lewenkopf, J. Phys. Cond. Matt. 22, 273201 (2010).

[31] S. Das Sarma, S. Adam, E. H. Hwang, and E. Rossi. Rev. Mod. Phys. 83, 407–470 (2011).

[32] S. Roche, N. Leconte, F. Ortmann, A. Lherbier, D. Soriano, and J. C. Charlier. Solid State Commun. 152, 1404–1410 (2012).

[33] A. Cresti, N. Nemec, B. Biel, G. Niebler, F. Triozon, G. Cuniberti, and S. Roche, Nano Res. 1, 361–394 (2008).

[34] J. Wurm, A. Rycerz, I. Adagideli, M. Wimmer, K. Richter, and H. U. Baranger, Phys. Rev. Lett. 102, 056806 (2009).

[35] J. Wurm, K. Richter, and I. Adagideli, Phys. Rev. B 84, 205421 (2011).

[36] A. Rycerz, Phys. Rev. B 85, 245424 (2012).

[37] M. Wimmer, A. R. Akhmerov, and F. Guinea, Phys. Rev. B 82, 045409 (2010).

[38] F. Libisch, C. Stampfer, and J. Burgdorfer, Phys. Rev. B 79, 115423 (2009).

[39] G. Pal, W. Apel, and L. Schweitzer, Phys. Rev. B 84, 075446 (2011).

[40] E. H. Lieb, Phys. Rev. Lett. 62, 1201–1204 (1989).